%% file: 02diff01.tex
\documentclass[11pt]{article}

\newtheorem{lemma}{Lemma}
\title{Corrections to Fluid Dynamics}
\author{R. F. Streater,\\Dept. of Mathematics, King's College, London}
\date{14 Oct 2002}
\begin{document}
\maketitle
\setlength{\oddsidemargin}{0in}
\setlength{\evensidemargin}{0in}
\setlength{\topmargin}{0in}
\begin{abstract}
We show that a Galilean invariant version of fluid dynamics can be
derived by the methods of statistical dynamics using Maxwell's
balance equations. The basic equation is non-local, and might
replace the Boltzmann equation if the latter turns out not to have global
smooth solutions in general. As an approximation, a local form of the
equations of motion is derived. It turns out to be a version of the
Navier-Stokes system, obeying the Stokes relation, and with the
viscosity
coefficient rising as $\Theta^{1/2}$ with temperature $\Theta$. The new
feature is the presence of the Dufour effect for a gas of a single
component. This ensures that the principal symbol of the parabolic system
is non-singular.
\end{abstract}

\section{Introduction}
A central problem for mathematical fluid dynamics is the
derivation of the Navier-Stokes equations ({\em N-S})
starting from a reversible dynamical theory such as classical hard
spheres or quantum mechanics. To achieve this, it seems essential
to know that the {\em N-S} equations themselves possess smooth solutions
for all times, for a large enough class of smooth initial values
for the fields. Hence the latter problem is revealed as the key
question, for one version of which a Clay Millenium prizes is offered.
The prize version describes an incompressible liquid
under isothermal conditions. Nash
\cite{Nash1,Nash2} had shown that, given smooth initial conditions, there
exists a unique smooth solution for a small enough time. The question
of smooth global-in-time solutions remains open except for small
initial conditions. To model driven systems such as B\'{e}nard convection,
the isothermal condition must be relaxed; recent numerical studies
\cite{Hoyas} show qualitative agreement with experiment. In \cite{Hoyas},
the condition of incompressiblitiy, div$\,${\boldmath$u$}=0 is maintained,
and the energy equation is modified by the addition of a bouyancy condition.
This is expressed by requiring that the liquid in hot regions is less
dense than in cold regions, a rather {\em ad hoc} procedure.
We shall argue in Sect.~(2) that {\em N-S} is at the boundary of a more
regular class of models. In {\em N-S}, the pressure is infinite, but its
place is taken by a surrogate pressure determined by the requirement of
self-consistency. Thus in {\em N-S}, the pressure is a balancing item
like petty cash, much beloved by accountants,
which can be adjusted to cancel errors made elsewhere in the calculation.

In this paper, {\em C-N-S-T} will denote the system of five coupled non-linear
partial differential equations known \cite{Lions} as `compressible
Navier-Stokes with temperature'. This might or might not be an easier
problem than {\em N-S}, but it is certainly more widely applicable.
A difficulty with {\em C-N-S-T} is that the symbol of the elliptic operator
is singular. To derive {\em N-S}, some authors
divide the problem into two parts \cite{Beck}; first,
to show that reversible dynamics is well represented in some limit
by a stochastic process; then to show that the Fokker-Planck
equations of this process gives rise to the {\em N-S} equations. The
latter is only partially achieved in \cite{Beck}. Other authors
{\em start} with some version of stochastic dynamics
\cite{Vishik,Manin,Wojnar}, and prove things; this approach will be
adopted here, as it avoids the much harder first part.
We shall adapt information geometry \cite{Ingarden1,Balian1,Kossakowski}
to the dynamics of a rare gas. Thus, the state of the system is not a point
in phase space, but a measure $\mu$ on it, and the dynamics is a path in
the space $\Sigma$ of measures.
In the usual treatment \cite{Balescu} the currents of the
conserved variables are given exactly by expressions involving higher
moments of the same random fields; the
time derivatives of these higher moments involve yet higher moments.
The whole system goes on for ever, and is called the {\em BBGKY} hierarchy.
Inasmuch as the system is equivalent to classical mechanics, it is
reversible and shows no dissipation. Artfulness is needed to `close' the
system in terms of the slow fields; that is, to write the time-derivatives
of the slow fields in terms of themselves, by truncating the system. It has
proved possible to get a variety of kinetic equations from the hierarchy, by
taking a limit appropriate to the physical situation \cite{Balescu}.
These limiting systems exhibit dissipation. It then remains to show that
the solutions to the limiting system are limits of solutions to the {\em
BBGKY} hierarchy. This programme has proved to be difficult to complete.

Information dynamics offers an alternative. In the case studied here,
where the potential between the
particles is zero outside the hard core, the states in {\em local
thermodynamic equilibrium, LTE} can be computed. The {\em LTE} states
are products over the lattice; the state at a site {\boldmath$x$} is of the
form
\[
\mu(\omega_{\mbox{\boldmath$x$}})=N_{\mbox{\boldmath$x$}}
p(\mbox{\boldmath$x,k$}),\]
where $p$ is Maxwellian.
The set of such states make up the information manifold, ${\cal M}$.
Any state $\mu\in\Sigma$, having finite means for the slow variables,
has a reduced description, denoted $\mu Q$;
here $Q$ is the the non-linear projection onto ${\cal M}$, acting on the
right, which maps $\mu$ to the state in ${\cal M}$ with the same means for
the slow variables as $\mu$. The Gibbs principle \cite{Jaynes,Ingarden2}
states that $\mu Q$ is the state of maximum entropy having these means.
In the original formulation of information
dynamics \cite{Ingarden1}, in the time interval $(0,t)$ the state $\mu$
evolves under the reversible dynamics of classical mechanics, to $\mu(t)$
say. This state has a much simpler description by the {\em LTE} state
$\mu(t)Q$, which cannot be distinguished from $\mu(t)$ by measuring the slow
variables. The orbit $\{\mu(t)Q:t\geq 0\}$ in ${\cal M}$ was intended as
the thermodynamic evolution. It is clear that the entropy of
$\mu(t)Q$ is not less than that of $\mu$; there is a transfer of information
into inaccessible degrees of freedom by the reversible motion \cite{Balian2}.
It is not always true that entropy increases along the orbit, as is seen
if the classical motion were periodic. Another version of information
dynamics was adopted in
\cite{Balian1}; there, the reversible motion took place for a very small
time $t$, and the reduced description $\mu(t)Q$ was used, instead of
$\mu(t)$, as the initial state of the next step. This gives a discrete-time
semigroup, with increasing entropy; however the time-step cannot be taken
to zero, without sending the rate of entropy production to zero as well
\cite{Kossakowski,Balian1}. It is necessary to keep the time-step
positive; it represents the relaxation time, and the map $Q$
implements the thermalisation of the state $\mu(t)$. The challenge is to
do this in a way that is invariant under the Galilean group (denoted
by ${\cal G}$ below).

Information dynamics has
been extended \cite{RFS1} to allow stochastic dynamics; then the time-step
can go to zero, still giving a non-zero rate of entropy production. Another
idea is to allow state-dependent transition rates \cite{Alicki}.
With these changes, one may
call the theory {\em statistical dynamics}. It is designed to obey both the
first and the second laws of thermodynamics, but otherwise puts
few constraints on the form of the dynamical equations. The choice made
for the dynamics determines the nature of the system under discussion.
At first sight, statistical dynamics has too much noise; a simple
application is shown \cite{RFS2} to lead to mass diffusion and
the Soret effect for an inert gas at rest, contrary to the
literature \cite{Balian2}. Indeed, without a velocity field, the
theory `has not got off the ground' \cite{Leppington}. Truesdell
\cite{Truesdell} ironically says ``results of this kind are
described by kinetic theorists as `corrections to hydrodynamics' ''.

In this paper, we apply statistical dynamics to the case of an
inert gas of a single type. We arrive at {\em C-N-S-T}, but with
one extra term, a Dufour effect. Thus the intuitively
attractive `method of Maxwell', \cite{Maxwell} in which we compute the
gain and loss of particles in a small time interval at each point
{\boldmath$x$}, is successful. The new idea is to postulate that in the
state $\mu$, some but
not all the particles are thermalised; those that are,
are described by the {\em LTE} state $\overline{\mu}$. However,
this state is NOT equal to $\mu Q$!
By construction, our model is the
the continuum limit of a non-linear Markov process on a lattice,
with a bistochastic transition matrix. This might be a possible
starting point for a proof that smooth solutions exist. The method of
Maxwell is treated in \cite{Chapman}, p 93, and in Balian \cite{Balian2},
but it is abandoned as too primitive, in favour of the Boltzmann equation.
This might have been too hasty, especially if the {\em BE} turns out not
to have smooth global solutions.

In Sect.~(2) we study a discrete model of hard spheres, for which
the equilibrium state exactly factorises. Sect.~(3) contains a
discussion of the idea that the true state $\mu$ should be
distinguished from its thermalised part, $\overline{\mu}$, using the
analogy of aircraft in an airfreight company. The hopping rules of
the nonlinear Markov chain are presented, and related to the mean
free time. We also suggest a ${\cal G}$-invariant collision
function. From this, we get an explicit formula
(\ref{meanfreetime3}) for the mean free time, $t_\ell$. In
Sect~(4) we start with the fundamental relation
(\ref{thenewnavier}) expressing the full state as a non-local
integral of the thermalised state. The dynamics is expressed in
terms of the {\em BBGKY} hierarchy. We obtain the Euler equations as the
zero$^{\rm th}$ approximation, and obtain a useful short version
of these. In Sect.~(5) we find the differences of the means of the
slow variables in the states $\mu$ and $\overline{\mu}$, and show how
the method can be extended to other variables. This enables us to
compute the {\em BBGKY} moments in terms of the means in the full
state and thus arrive at a version of {\em C-N-S-T}. This exhibits
the Dufour effect, contrary to the literature \cite{Balian2}.

\section{The Thermostatics of an Inert Gas}
We take space to be $\Lambda\subseteq(a{\bf Z})^3$, and suppose the length
$a$, representing the diameter of a molecule, to be so small compared with
the variation of the macroscopic fields
that we can replace all sums over $\Lambda$ by integrals. The possible
configurations of the fluid are the points in the product sample space
\[
\Omega=\prod_{\mbox{\boldmath$x$}\in\Lambda}\Omega_{\mbox{\boldmath$x$}},\]
so a configuration is specified by the collection
$\{\omega_{\mbox{\boldmath$x$}}\}_{\mbox{\boldmath$x$}\in\Lambda}$.
For each {\boldmath$x$},
\[
\Omega_{\mbox{\boldmath$x$}}=\left\{\emptyset,(\epsilon{\bf
Z})^3\right\}.\] Here, $\epsilon$ is a small parameter having the
dimension of momentum; for example, we could take $a\epsilon=h$,
the semi-classical division of the phase-space of a particle into
cubes of volume $h^3$. If the system is in a configuration
$\omega$, such that $\omega_{\mbox{\boldmath$x$}}=\emptyset$, then
we say that the site {\boldmath$x$} is empty. If
$\omega_{\mbox{\boldmath$x$}}=\mbox{\boldmath$k$}$, we say that
the site {\boldmath$x$} is occupied, by a particle of momentum
{\boldmath$k$}. This simple exclusion of more than one particle on
each site incorporates the hard-core repulsion between the
particles, which are thus hard spheres sitting at some of
the points of $\Lambda$. The field point of view enables us to
avoid the Gibbs paradox.

The {\em state} of the system is a probability on $\Omega$,
denoted by $\mu$. We denote the set of states by
$\Sigma$. The `slow variables' of our model
are the 5 extensive conserved random fields
\begin{eqnarray}
{\cal N}_{\mbox{\boldmath$x$}}(\omega)&=&\left\{\begin{array}{ll}
               0&\mbox{ if }\omega_{\mbox{\boldmath$x$}}=\emptyset\\
               1&\mbox{ if }\omega_{\mbox{\boldmath$x$}}=\mbox{\boldmath$k$}
               \end{array}\right.\\
{\cal E}_{\mbox{\boldmath$x$}}(\omega)&=&\left\{\begin{array}{ll}
               0&\mbox{ if }\omega_{\mbox{\boldmath$x$}}=\emptyset\\
               \mbox{\boldmath$k\cdot k$}/2m+\Phi(\mbox{\boldmath$x$})
&\mbox{ if }\omega_{\mbox{\boldmath$x$}}=\mbox{\boldmath$k$}
               \end{array}\right.\\
\mbox{\boldmath${\cal P}_x$}(\omega)&=&\left\{\begin{array}{ll}
               0&\mbox{ if }\omega_{\mbox{\boldmath$x$}}=\emptyset\\
               \mbox{\boldmath$k$}&\mbox{ if
               }\omega_{\mbox{\boldmath$x$}}
=\mbox{\boldmath$k$}
               \end{array}\right.
\end{eqnarray}
Here, $\Phi(\mbox{\boldmath$x$})$ is the external potential energy per
particle. The variables appearing in the {\em C-N-S-T} equations are simply
related to the {\em  mean fields} in the state $\mu$:
\begin{equation}
N_{\mbox{\boldmath$x$}}={\bf E}_\mu[{\cal N}_{\mbox{\boldmath$x$}}];
\hspace{.4in}E_{\mbox{\boldmath$x$}}={\bf E}_\mu[{\cal
E}_{\mbox{\boldmath$x$}}];\hspace{.4in}
\mbox{\boldmath$\Pi_x$}={\bf E}_\mu[\mbox {\boldmath${\cal
P}_x$}]. \label{means}
\end{equation}
In information geometry, the specification of the slow variables
determines the {\em information manifold} ${\cal M}$, which in the
context of fluid dynamics consists of states in {\em LTE} (local
thermodynamic equilibrium). Such a state is specified by five {\em
canonical fields}, dual to the mean fields:
$\beta_{\mbox{\boldmath$x$}},\xi_{\mbox{\boldmath$x$}},\mbox{\boldmath$\zeta
_x$}$, and has the form
\begin{equation}
\mu(\omega)=\prod_{\mbox{\boldmath$x$}\in\Lambda}\Xi_{\mbox{\boldmath$x$}}
^{-1}\exp\left\{-\xi_{\mbox{\boldmath$x$}}{\cal N}_
{\mbox{\boldmath$x$}}(\omega)-\beta_{\mbox{\boldmath$x$}}{\cal
E}_{\mbox{\boldmath$x$}}(\omega)-
\mbox{\boldmath$\zeta_x\cdot{\cal P}_x$}(\omega)\right\}.
\label{LTE}
\end{equation}
In finding the partition function
\begin{equation}
\Xi_{\mbox{\boldmath$x$}}=1+\epsilon^{-3}\left(\frac{2\pi
m}{\beta_{\mbox{\boldmath$x$}}}\right)^{3/2}\exp
\left\{-\xi_{\mbox{\boldmath$
x$}}-\beta_{\mbox{\boldmath$x$}}\Phi(\mbox{\boldmath$x$})+m\mbox{
\boldmath$\zeta_x\cdot\zeta_x$}/2\beta_{\mbox{\boldmath$x$}}\right\}
\end{equation}
we have replaced the sum over the momentum lattice of size
$\epsilon$ by a Gaussian integral. The product structure of an
{\em LTE} state means that an observable at a point of $\Lambda$
is independent of an observable at any other. The state
$\mu$ can be written in Maxwell form
\begin{equation}\label{Maxwell}
\mu=N_{\mbox{\boldmath$x$}}p(\mbox{\boldmath$x,k$})=N_{\mbox{\boldmath$x$}}Z^{-1}\exp\left\{
-\beta_{\mbox{\boldmath$x$}}\Phi(\mbox{\boldmath$x$})-
\beta_{\mbox{\boldmath$x$}}\mbox{\boldmath$k\cdot
k$}/(2m)-\mbox{\boldmath$\zeta_x\cdot
k$}\right\},
\end{equation}
where
\begin{equation}
Z_x=\epsilon^{-3} \left(\frac{2\pi
m}{\beta_{\mbox{\boldmath$x$}}}\right)^{3/2} \exp\left\{-\beta_
{\mbox{\boldmath$x$}}\Phi(\mbox{\boldmath$x$})+ \frac{m\mbox
{\boldmath$\zeta_x\cdot\zeta_x$}}
{2\beta_{\mbox{\boldmath$x$}}}\right\}. \label{Z}
\end{equation}
We note the identity for each {\boldmath$x$}
\[\Xi=1+e^{-\xi}Z.\]
The external potential does not influence the local velocity
distribution, as it is cancelled out by the partition function.
 The mean fields (\ref{means}) are related to the canonical fields
by
\begin{eqnarray}
E_{\mbox{\boldmath$x$}}=-\frac{\partial}{\partial
\beta_{\mbox{\boldmath$x$}}}
\log\Xi_{\mbox{\boldmath$x$}}&=&N(\mbox{\boldmath$x$})
\left(\Phi(\mbox{\boldmath$x$})+\frac{3}
{2\beta_{\mbox{\boldmath$x$}}}+\frac{m\mbox{\boldmath$\zeta_x\cdot\zeta_x$}}
{2\beta_{\mbox{\boldmath$x$}}^2}\right)\label{ene}\\
N_{\mbox{\boldmath$x$}}=-\frac{\partial}{\partial\xi_{\mbox{\boldmath$x$}}}
\log\Xi_{\mbox{\boldmath$x$}}&=&\frac{\Xi_{\mbox{\boldmath$x$}}-1}
{\Xi_{\mbox{\boldmath$x$}}}=\frac{Ze^{-\xi_{\mbox{\boldmath$x$}}}}
{1+Ze^{-\xi_{\mbox{\boldmath$x$}}}}\label{num}\\
\Pi_{\mbox{\boldmath$x$}}^i=-\frac{\partial}{\partial\zeta_i}\log\Xi_{\mbox{\boldmath$x$}}
&=&-\frac{mN_{\mbox{\boldmath$x$}}\zeta_{\mbox{\boldmath$x$}}^i}
{\beta_{\mbox{\boldmath$x$}}}.\label{mom}
\end{eqnarray}
The formalism breaks down if $\beta$ is zero or infinity, or if
$N$ vanishes, but the case of a fluid at rest,
$\mbox{\boldmath$\Pi$}=0$, is within the information manifold, ${\cal
M}$.

Historically, the intensive variables used in the N-S equations
were the chemical potential $-\xi/\beta$, the velocity field
$\mbox{\boldmath$u$}=-\mbox{\boldmath$\zeta$}/\beta$ and the
temperature $\Theta=\left(k_{_B}\beta\right)^{-1}$. We shall
eliminate $\xi$ in favour of the mass-density $\rho=a^{-3}mN$
using (\ref{num}), which leads to
\begin{equation}
e^{-\xi_{\mbox{\boldmath$x$}}}=Z_{\mbox{\boldmath$x$}}^{-1}
N_{\mbox{\boldmath$x$}}/\left(1-N_{\mbox{\boldmath$x$}}\right).
\end{equation}
The mean occupation per site $N_{\mbox{\boldmath$x$}}$ obeys
$0<N_{\mbox{\boldmath$x$}}<1$. The (von Neumann) entropy of any
state $\mu$ is
\begin{equation}\label{entropy}
S(\mu):=-k_{_B}\sum_\omega\mu(\omega)\log\mu(\omega).
\end{equation}
Gibbs knew that the state of maximum entropy, among all states
with the given means of the total energy and number of particles,
is the exponential state that he called the grand canonical state
\cite{Jaynes,Ingarden2,Kossakowski}. This is a simple exercise in Lagrange
multipliers. If the mean fields depend on {\boldmath$x$}, then the
state of maximum entropy has the same form, in which the canonical
fields $\xi,\beta$ and $\zeta$ now depend on {\boldmath$x$}.

In this section we study the system in equilibrium,
and denote by $E,N$ and {\boldmath$\Pi$} the total values of
the energy, number and momentum; then (\ref{entropy}) gives for
the entropy
\begin{equation}
\Theta S(\mu)=E+k_{_B}\Theta\xi N-\mbox{\boldmath$u\cdot\Pi$}+
k_{_B}\Theta\log\Xi.
\end{equation}
Compare this with the thermostatic formula
\begin{equation}
\Theta S=E+k_{_B}\Theta\xi N-\mbox{\boldmath$u\cdot\Pi$}+PV
\end{equation}
(note that the term {\boldmath$u\cdot\Pi$} is omitted in
\cite{Lions}, eq.~(1.17)), where $P$ is the pressure and $V$ is
the volume; we see that
\begin{equation}
P=k_{_B}\frac{|\Lambda|}{V}\Theta\log\Xi=k_{_B}\Theta
a^{-3}\log\Xi.
\end{equation}
If there are $N=\sum_{\mbox{\boldmath$x$}}N_{\mbox{\boldmath$x$}}$
particles, and $V_0$ is the smallest volume they can occupy (one
per site), then $V_0=a^3N$ and $N_{\mbox{\boldmath$x$}}=V_0/V$.
Also,
\[
\Xi_{\mbox{\boldmath$x$}}=(1-N_{\mbox{\boldmath$x$}})^{-1}=1+V_0/(V-V_0).\]
Thus at equilibrium, we have the equation of state
\begin{equation}
P=\frac{k_{_B}\Theta}{V_0}N\log\left(1+\frac{V_0}{V-V_0}\right).
\label{pressure}
\end{equation}
For small $V_0/V$ this is close to the van der Waals gas
\begin{equation}
(P+A/V^2)(V-V_0)=Nk_{_B}\Theta
\label{Waals}
\end{equation}
with $A=0$. Unlike the case $A>0$, this model shows no failure in
convexity in its isothermals.

\input 02diff02

\end{document}

%% file: 02diff02.tex
\section{The Statistical Dynamics of the Gas}
\subsection{An Airfreight Model}
Consider a gas of free particles in a box with reflecting walls, in
equilibrium; then
{\boldmath$u$}, $\Theta$ and $\rho$ do not depend on
{\boldmath$x$}. There is still a lot going on. In a volume $d^3x$
around {\boldmath$x$}, a particle of momentum {\boldmath$k$},
which is present with probability $Np(\mbox{\boldmath$k$})$, moves
in the direction of the unit vector $\hat{\mbox{\boldmath$k$}}$,
to be replaced in time $t$ by a particle with the same
{\boldmath$k$} arriving from the point $\mbox{\boldmath$x-k$}t/m$.
This replacement was present with exactly the same probability.
The larger {\boldmath$k$} is, the further away is the source of
the replacement. In this picture, equilibrium is described by a
huge game of musical chairs; only the indistinguishability of the
particles
prevents this from being detected. Now look at the same mechanism,
but where $\rho$, \mbox{\boldmath$\Pi$} and $\Theta$ depend on {\boldmath$x$}.
That is, we now consider the Knudsen gas. There is no longer exact
replacement of the lost particles at
{\boldmath$x$}; the parameters $\rho\ldots$ change with time. It
might seem that the system gets closer to equilibrium, since the
parameters start to become more and more nearly constant. Of
course, the entropy is constant in time, as the system is
Hamiltonian, (free, even). The apparent increase in entropy
associated with the slow variables is exactly matched by a
reduction of entropy in inaccessible observables
\cite{Balian1,Balian2}. Thus, taking the initial state to be in
{\em LTE}, the random variables ${\cal N}\ldots$ had independent
values at every point. But after some time, the free motion
introduces correlations between very far points; if {\boldmath$x$}
at time $t$ has a particle with momentum {\boldmath$k$}, then
$\mbox{\boldmath$x-k$}t/m$ must have had a particle of momentum
{\boldmath$k$} at time $0$; so it did not have a particle of
momentum $\mbox{\boldmath$k$}^\prime\neq\mbox{\boldmath$k$}$. Thus
$\mbox{\boldmath$x$}+(\mbox{\boldmath$k$}^\prime-\mbox{\boldmath$k$})
t/m$ has no particle of momentum $\mbox{\boldmath$k$}^\prime$ at time
$t$, a statement correlated with the assumption, above, about
{\boldmath$x$} at time $t$. Correlations like these might at any
time show up in behaviour quite unlike that of a system near
equilibrium. For spin systems, the spin-echo effect is such an
example \cite{Balian2}.
In a real gas, we do not expect any surprises such as the spin-echo
effect. This is due to the interactions, idealised by collisions, which
remove correlations between distant points, and also help to redistribute
energy and momentum. Collisions do not contribute to transport; on the
contrary, they inhibit the free flow: the diffusion constant is inversely
proportional to the collision cross-section.

The concept of whether a system is thermalised or not is
independent of the Galilean frame of reference used in the
description. This is expressed mathematically by the fact that the
set of equilibrium states is mapped to itself by the group ${\cal G}$. The set ${\cal M}$ of states in {\em LTE} is
also mapped to itself by ${\cal G}$, which we interpret as saying
that the concept of partially thermalised systems is also
invariant under ${\cal G}$. Physically, a gas consists of some
(most) particles that are thermalised, and are described by an
{\em LTE} state with means equal to the averages over the
thermalised particles. A smaller number are not well described in
this way; in particular, their correlations with other particles
are underestimated by assuming {\em LTE}. Moreover, a particle
that is thermalised at time $t$ will move under its free motion to
regions at different density and temperature, and so after some
time (how short depends on the gradients) it will not be well
described as being thermalised. On the other hand, particles left
out of the count of thermalised particles make collisions during
their relaxation time, and return to the thermal state. The mass,
energy and momentum will on average be conserved, but a particle
can leave the {\em LTE} state at one point, and another can return
at another point, so the microscopic currents describing all
particles may differ slightly from the currents of thermalised
particles as described by the {\em LTE} state. This division of
the state into thermalised particles and the rest differs from the
division presented e.g. in \cite{Balescu}, p. 160, where the {\em
LTE} part of the state gives the same expectation values as the
true state. On the contrary, in our division, the state
$\overline{\mu}$ does not give the same means as the true state $\mu$.
A different division can be found in \cite{Grad}, p 229,
where $nF_1^\sigma$ is taken to be `the density of particles that
are not undergoing collisions at the given instant'. From now on,
the variables $N,E,\mbox{\boldmath$\Pi$}$, etc.,  refer to the
full state, and written with bars, they refer to the {\em LTE}
state of the thermalised particles.

The dynamics of a classical gas of hard spheres is similar to that
of an airfreight company, whose planes fly between airstrips
arranged in a lattice $\Lambda$. In calculating the overall
transport of goods, the company uses statistical methods; they
have records only of the local averages, at each airstrip,  of the
number of planes, their velocities, and their kinetic energies, at
time $t=0$. Every plane is instructed to select a velocity from
the Maxwell distribution at its airstrip, and to fly with this
velocity until it meets another plane in its airspace. Both planes
must then land very briefly, and record their presence, momentum
and energy to the local computer. This recalculates the updated
values of $N, \mbox{\boldmath$\Pi$},E$ for this strip, and
instructs them to take off with new velocities drawn from the
updated Maxwellian. The problem is to find a theory which can
predict the average transport of goods, ${\cal N},
\mbox{\boldmath${\cal P}$},{\cal E}$, without reading the local
computers. To account for the transport of
$N,\mbox{\boldmath$\Pi$},E$, we must introduce accounting system
non-local in space and time: we know that just after a landing and
take-off, the distribution of velocities is Maxwellian. This
simple fact leads us to the fundamental equation
(\ref{thenewnavier}).

We want the dynamics to satisfy the second law of thermodynamics.
This is ensured in a Markov chain if the transition matrix is
bistochastic. Physically, if the gas has no velocity
{\boldmath$u$}, the transition rate from {\boldmath$x$} to
{\boldmath$y$} by a particle of momentum {\boldmath$k$} is the
same as the rate from {\boldmath$y$} to {\boldmath$x$} by a
particle of momentum {\boldmath$-k$}. This expresses {\em
time-reversal invariance} of the transition matrix. That is, if
$\tau:\omega\mapsto \omega\tau$ is the time-reversal map on the
sample space, we say a transition matrix $T$ obeys time reversal
symmetry if
\begin{equation}
T(\omega|\omega^\prime)=T(\omega^\prime\tau|\omega\tau)\hspace{.5in}
\mbox{ for all }\omega,\omega^\prime\in\Omega.
\label{Tinv}
\end{equation}
We note the following lemma, whose proof is simple:
\begin{lemma}
Let $T$ be a stochastic map obeying (\ref{Tinv}); then $T$ is
bistochastic.
\end{lemma}
Our model dynamics will be given by a stochastic map
obeying (\ref{Tinv}), and so, by the lemma, will be bistochastic,
and so entropy-increasing.

\input 02diff03

%% file: 02diff03.tex

\subsection{The Hopping Rules} In this section, we give hopping
rules for the case of zero external field, $\Phi=0$. The dynamics
will be invariant under ${\cal G}$. We start with a model in
discrete space-time. The discrete dynamics will be given by
specifying a hopping probability in one time-step. In the
classical hard-sphere model, it is to be expected that on average,
particles of different speed travel through the same amount of
material before thermalising. Let $\ell(\mbox{\boldmath$x,k$})$
denote the average distance travelled by a particle starting at
{\boldmath$x$} with momentum {\boldmath$k$}; to begin with, assume
that {\boldmath$k$} lies along one of the basis vectors of the
lattice. $\ell$ is called the {\em mean free path}, and it
generalises an idea going back to Clausius \cite{Clausius}; it is
going to be the mean of a random distance $r$, the free path
between collisions. We assume, as part of the model, that the
particle thermalises on its first collision. The relaxation time
$t$ of a particular particle depends on its speed; for a particle
travelling the free path $r$, $t$ is the time taken,
$rm/|\mbox{\boldmath$k$}|$. We therefore cannot choose a unique
time-interval for the time step of all processes, and it seems
difficult to construct a discrete-time stochastic process. We
overcome this complication by noting the {\em rate}
$|\mbox{\boldmath$k$}|/(rm)$ at which the process transfers mass,
energy and momentum; we can then move to a continuous time process
with the same rate. We shall work with $t$
and its mean, $t_\ell$, rather than with the free path $r$ and its
mean, $\ell$; $t_\ell$ has the advantage of being the same in all
inertial frames.

The dynamics must be such as to conserve the totals
\[
{\cal N}:=\sum_{\mbox{\boldmath$x$}\in\Lambda}{\cal
N}_{\mbox{\boldmath$x$}} ;\hspace{.5in}{\cal
E}:=\sum_{\mbox{\boldmath$x$}\in\Lambda}{\cal
E}_{\mbox{\boldmath$x$}};\hspace{.5in}\mbox{\boldmath${\cal P}$}:=
\sum_{x\in\Lambda}\mbox{\boldmath${\cal P}_x$}.\]
These random variables divide
$\Omega$ into simultaneous level sets, the mass-shells,
energy-shells, and momentum shells, thus:
\[
\Omega_{N,E,\mbox{\boldmath$\Pi$}}:=\left\{\rule{0cm}{.6cm}\omega\in
\Omega:{\cal N}(\omega)=N,{\cal E}(\omega)=E,\mbox{\boldmath${\cal
P}$}(\omega)=\mbox{\boldmath$\Pi$}\right\}.\] 
Clearly,
\[
\Omega=\bigsqcup\Omega_{N,E,\mbox{\boldmath$\Pi$}}.\]
The dynamics, the Markov matrix $T$, must be chosen so that a point
$\omega$ jumps to another point in
the same shell. We cannot expect this to be given by a symmetric Markov
transition matrix: the inverse process involves a change of sign for
{\boldmath$k$}; however, we shall be able to construct a suitable
bistochastic map.
In fact, our Markov transition matrix $T$ will be a convex mixture
of permutations that
move a configuration $\omega_1$ to $\omega_2$, where in
$\omega_1$ there is a particle at {\boldmath$x$} and a hole at
{\boldmath$y$}, and in $\omega_2$ the opposite holds, with the mass, energy
and momentum that was at {\boldmath$x$} transported to {\boldmath$y$}.
This move is only possible if
all the points between {\boldmath$x$} and {\boldmath$y$} are empty.
Moreover, to ensure that at the end of the flight the particle returns to
the thermalised fold, the site one place past {\boldmath$y$} must be
occupied. We postulate that a particle moves in a straight line
along empty sites until it meets a filled site;
it then thermalises at the last empty site, {\boldmath$y$} say, and dumps
its mass, energy and momentum there, which
joins the mass, energy and momentum of the state $\mu_{\mbox{\boldmath$y$}}$.
We postulate one such jump for each $r=sa$, where $s$ is an integer, and for
each momentum {\boldmath$k$} at {\boldmath$x$}, and then for each
{\boldmath$x$}.

Suppose that the site {\boldmath$x$} is occupied, with momentum pointing
in the direction of one of the basis vectors of the lattice, say
{\boldmath$k=|k|e$}. The probability that the sites
$\mbox{\boldmath$x$}+s^\prime a\mbox
{\boldmath$e$}$ be empty, $1\leq s^\prime\leq s$, and
the site $\mbox{\boldmath$x$}+(s+1)a\mbox{\boldmath$e$}$
occupied, is
\begin{equation}
\prod_{s^\prime=0}^s\left(1-N_{\mbox
{\boldmath$x$}+s^\prime a\mbox{\boldmath$e$}}\right)
N_{\mbox{\boldmath$x$}+(s+1)a\mbox{\boldmath$e$}}
\label{firsttry}
\end{equation}
One can check that e.g. if $N_{\mbox{\boldmath$y$}}>0$ is
independent of {\boldmath$y$} for large enough $|\mbox
{\boldmath$x-y$}|$, then
\begin{equation}
\sum_{s}\prod_{s^\prime=0}^s\left(1-N_{\mbox
{\boldmath$x$}+s^\prime a\mbox{\boldmath$e$}}\right)N_{\mbox{\boldmath$x$}+(s+1)a
\mbox{\boldmath$e$}}=1.
\label{sumisone}
\end{equation}
This just expresses that with probability one, the number of holes
adjacent to {\boldmath$x$} on the line joining {\boldmath$x$} to
infinity along the direction {\boldmath$e$} must be some integer.
The product in (\ref{firsttry}) is a marginal probability of the
state $\mu$, and so is linear in $\mu$; however, it is a polynomial of
indefinite degree in the variables $N_{\mbox{\boldmath$x$}}$, in
terms of which the equations of motion are to be written. In
discussing the flow of mass, energy and momentum at the point
{\boldmath$x$}, it is convenient to include this factor in the
hopping probability, rather than in the initial state. We then get
a Markov chain on the probability space of the two points at the
ends of the path,
\begin{equation}
\Omega_{\mbox{\boldmath$x$}}\times\Omega_{\mbox{\boldmath$y$}};
\label{twopoint}
\end{equation}
the transition probability depends on the state
$\mu$, but otherwise obeys the properties of a bistochastic map. This allows
the continuum limit of (\ref{firsttry}) to be taken without leaving
elementary probability theory. In this limit, we define the densities
\[\rho(\mbox{\boldmath$x$}):=
ma^{-3}N_{\mbox{\boldmath$x$}};\hspace{.6in}
\mbox{\boldmath$\varpi$}:=a^{-3}\mbox{\boldmath$\Pi$}_{\mbox{\boldmath$x$}},\]
and get a damped exponential
\[\prod_{s^\prime=0}^s\left(1-N{\mbox{\boldmath$x$}+s^\prime a\mbox
{\boldmath$e$}}\right)\sim
\exp\left\{-\frac{a^2}{m}\int_0^r\rho(\mbox{\boldmath$x$}+
r^\prime\mbox{\boldmath$e$})dr^\prime\right\},\hspace{.3in}a\rightarrow 0.\]
We shall use an identity similar to (\ref{sumisone}) to find the mean 
free time, subject to two refinements. First we require that the site
$\mbox{\boldmath$x$}+sa\hat{\mbox{\boldmath$k$}}$ be empty {\em at
the time $t(s^\prime)$ that this point is reached} by our
travelling particle. By the same argument, the free path is $r=sa$
if the site $\mbox{\boldmath$x$}+ a(s+1)\mbox{\boldmath$e$}$ is
occupied at the time at which the arriving particle reaches it.
This refinement leads to a transition rate is non-local in the time; it 
will turn out that in the model we construct, we can replace these
intermediate times by the current time with error of second order.
With this done, the transition probability is invariant under
time-reversal. It then follows from Lemma 1 that, by adjusting the
stay-put probability so that the rows add up to one, we get a
bistochastic map on the two-point sample space.

The second refinement comes from the requirement of 
${\cal G}$-invariance in the continuum limit. We
claim that the `thermalised part' of a state should be a ${\cal
G}$-invariant concept, and this will be achieved by (\ref{basic}).
The Euler dynamics, which is close to the true dynamics, brings
the state $\overline{\mu}$ out of {\em LTE} in any time interval.
The part of the state not thermalised soon becomes thermalised by
collisions, and it is this thermalisation that is involved in the
dissipative part of the dynamics. There is some ambiguity in the choice of
splitting, because {\em all} the particles leaving {\boldmath$x$}
seem to be instantly dethermalised unless $\mu$ is at equilibrium.
However, inasmuch as the gradients are small, some part of the
thermal state at time $t_0$ might remain thermalised at time
$dt_0$ later. How do we decide on how much? It is easy to agree
about the rate at which particles thermalise at {\boldmath$x$} at
time $t_0$. We include those particles arriving at {\boldmath$x$}
having a momentum {\boldmath$k$} and a relaxation time $t^\prime$, 
and originating at $\mbox{\boldmath$x$}-\mbox{\boldmath$k$}t^\prime/m$. 
The rate at which they transfer mass, momentum and energy is $1/t^\prime$
times the mass, momentum and energy they carry.

We note that the concept of thermalisation at a point
{\boldmath$x$} over a time-interval $(0,t)$ is not a ${\cal
G}$-invariant concept. For suppose that in the inertial frame
${\cal O}$, various particles thermalise at {\boldmath$x$} at
times $0<t_1<\ldots <t_n<t$; then in an inertial frame ${\cal
O}^\prime$, moving relative to ${\cal O}$ with velocity
{\boldmath$V$}, they will thermalise (at the same times) at the
points $\mbox{\boldmath$x$}+ \mbox{\boldmath$V$}t_i$. We therefore
must consider the thermalisation (and dethermalisation) that
occurs at $t$ in a time-interval $(t,t+dt)$, for an infinitesimal
$dt$; this has an invariant meaning. The number thermalising in
any time interval of length $dt$ is $dt$ times the rate at which
the thermalisation occurs.

The continuum analogue of (\ref{firsttry}) is a function $w(\mbox{\boldmath
$x,k$},t_0;t)$ which is the probability density that a particle at
{\boldmath$x$} with momentum {\boldmath$k$} will travel exactly a distance
$r=|\mbox{\boldmath$k$}|t/m$ and then thermalise at $\mbox{\boldmath$y=x$}+
\mbox{\boldmath$k$}t/m$ in the time interval $(t_0+t,t_0+t+dt)$.
Let $W$ denote the probability density that a particle at {\boldmath$x$}
with momentum {\boldmath$k$} has had no collision up to the point
{\boldmath$y$}, and let $C(\mbox{\boldmath$y,k$},t+t_0)$ (for collisions)
be the probability density that a particle at {\boldmath$y$} with momentum
{\boldmath$k$} at time $t_0+t$ will be thermalised in the tube of diameter
$a$ and length $dr=|\mbox{\boldmath$k$}|dt/m$. Then we have
\begin{equation}
w(\mbox{\boldmath$x,k$},t_0;t)=W(\mbox{\boldmath$x,k$},t_0;t)C(\mbox
{\boldmath$x$}+\mbox{\boldmath$k$}t/m,t_0+t).
\end{equation}
Then the analogue of (\ref{sumisone}) is
\begin{equation}
\int_0^\infty w(\mbox{\boldmath$x,k$},t_0;t)dt=1
\label{sumisone2}
\end{equation}
for all {\boldmath$x,k$} and $t_0$. The mean relaxation time is defined
to be
\begin{equation}
\int_0^\infty w(\mbox{\boldmath$x,k$},t_0;t)t\,dt=t_\ell(\mbox{\boldmath$x,k$},
t_0).
\label{relax}
\end{equation}
We note that (\ref{sumisone2})
\begin{equation}
\int_0^\infty W(\mbox{\boldmath$x,k$},t_0;t)C(\mbox{\boldmath$x$}+\mbox
{\boldmath$k$}t/m,\mbox{\boldmath$k$},t_0+t)dt=1
\label{sumisone3}
\end{equation}
can be solved in terms of $C$:
\begin{equation}
W(\mbox{\boldmath$x,k$},t_0;t)=\exp\left\{-\int_0^t
C(\mbox{\boldmath$x$}+
\mbox{\boldmath$k$}t_1/m,\mbox{\boldmath$k$},t_0+t_1)dt_1\right\}.
\label{W}
\end{equation}
To see (\ref{W}), differentiate to get
\[
\frac{\partial W}{\partial t}=-WC,\]
and this is just minus the integrand in (\ref{sumisone3}). Then one 
verifies (\ref{sumisone3}):
\[
\int_0^\infty W(\mbox{\boldmath$x,k$},t_0;t)C(\mbox{\boldmath$x,k$},t_0+t)=
-\int_0^\infty\frac{\partial W}{\partial t}dt=-(W(\infty)-W(0))=1\]
assuming that $\rho$ is bounded away from zero along the line 
$\mbox{\boldmath$x$}+\mbox{\boldmath$k$}t/m$; this is enough to ensure that the mean free path is
finite.
By construction, $w=-\partial_t W$ and another form for the mean
free time is
\begin{equation}
t_\ell(\mbox{\boldmath$x,k$},t_0)=\int_0^\infty W(\mbox{\boldmath$x,k$},t_0;t)
\,dt.
\label{tell}
\end{equation}

\subsection{The Collision Function}
We now find the collision function $C$ appropriate for a hard-core
gas. In the dynamics of the lattice model, particles hop from site
to site with various rates. To implement invariance under ${\cal
G}$ , we extend the fields $\rho,\,E,\,\mbox{\boldmath$\varpi$}$
from $\Lambda$ to its convex hull ${\bf R}^3$ as continuous
piecewise linear functions, and consider a particle at
{\boldmath$x$} with momentum {\boldmath$k$} not necessarily lying
along a basis vector. Although the path of such a particle might
not intersect $\Lambda$, we must assign a rotated version of $w$
as the probability of a free time $t$, such that (\ref{sumisone2})
holds. We can assume that the particle hops to the nearest site of
$\Lambda$ to {\boldmath$x$}$+${\boldmath$k$}$t/m$. The continuum
limit then makes sense, with densities replacing probabilities.
Now divide the event, `the free time is $t$' into the subevents
`the free time is $t$, and $A$ collides with a particle $B$ of
momentum {\boldmath$q$}'. Let $w(\mbox{\boldmath$x,k,q$},t)$ be
the probability density for this. Then
\begin{equation}
w(\mbox{\boldmath$x,k$},t)=\int d^3q\,w(\mbox{\boldmath$x,k,q$},t).
\end{equation}
If $t$ increases to $t+dt$, then the free path $r$ increases to $r+dr$,
where $dt=m\,dr/|\mbox{\boldmath$k$}|.$ The probability that $A$ meets $B$
must be independent of the Galilean frame of reference. Consider
the frame ${\cal O}^\prime$ in which $B$ is at rest. Let
\begin{eqnarray}
(\mbox{\boldmath$x$},t)&\mapsto&
(\mbox{\boldmath$x$}^\prime,t^\prime)
=(\mbox{\boldmath$x$}+\mbox{\boldmath$q$}t/m,t)\nonumber\\
\mbox{\boldmath$k$}&\mapsto&\mbox{\boldmath$k-q$}\label{G}
\end{eqnarray}
be the Galilean transformation, and denote by $\rho^\prime,
\mbox{\boldmath$u$}^\prime,\Theta^\prime, p^\prime$ the {\em C-N-S-T}
variables and the probability as observed in ${\cal O}^\prime$. Then
\begin{eqnarray}
\rho^\prime(\mbox{\boldmath$x$}^\prime,t^\prime)&=&\rho(\mbox{\boldmath$x$}
,t)\label{rhodash}\\
\Theta^\prime(\mbox{\boldmath$x$}^\prime,t^\prime)&=&\Theta(
\mbox{\boldmath$x$},t)\label{thetadash}\\
\mbox{\boldmath$u$}^\prime(\mbox{\boldmath$x$}^\prime,t^\prime)&=&
\mbox{\boldmath$u$}(\mbox{\boldmath$x$},t)-\mbox{\boldmath$q$}/m\label{udash}
\\
p^\prime(\mbox{\boldmath$y$}^\prime,\mbox{\boldmath$q$}
^\prime,t^\prime)&=&
p(\mbox{\boldmath$y,q$},t).\label{pdash}
\end{eqnarray}
Then $A$ has momentum {\boldmath$k-q$} and in time $dt$ (which is
the same in all Galilean frames) $A$ sweeps out a region of volume
$dV=\sigma|\mbox{\boldmath$k-q$}|dt/m$, where $\sigma$ is the cross-section.
It meets a particle in
this volume having momentum {\boldmath$0$} with probability
\[
m^{-1}dV\rho'(\mbox{\boldmath$y$}',t)p^\prime(\mbox{\boldmath$y$}
^\prime,\mbox{\boldmath$0$},t). \]
By invariance, this is also the probability of collision
in the original frame, which is therefore
\begin{equation}
\sigma|\mbox{\boldmath$k-q$}|(dt/m^2)\rho^\prime(\mbox{\boldmath$y$}^\prime,
t^\prime)p^\prime(\mbox{\boldmath$y$}^\prime,\mbox{\boldmath$0$},
t^\prime)=\sigma|\mbox{\boldmath$k-q$}|(dt/m^2)
\rho(\mbox{\boldmath$y$},t)p(\mbox{\boldmath$y,q$},t),
\end{equation}
by (\ref{rhodash}) and (\ref{pdash}). This suggests the choice of collision
term
\begin{equation}
C(\mbox{\boldmath$y,k$},t_0+t)=\frac{\sigma}{m^2}\int d^3q|\mbox
{\boldmath$k-q$}|\rho(\mbox{\boldmath$y$},t+t_0)p(\mbox{\boldmath
$y,q$},t_0+t).
\label{collisionterm}
\end{equation}
Note that we include collisions in which the particle $A$ is hit
from behind by the particle $B$.
\subsection{The mean free time}
Our formula (\ref{collisionterm}) for $C$ gives for the mean free time
\begin{eqnarray}
t_\ell(\mbox{\boldmath$x,k$})&=&\int_0^\infty
\,dt\exp\left\{-\frac{\sigma}{m^2}\int_0^t
dt_1\rho(\mbox{\boldmath$x$}+\mbox{\boldmath$k$}t_1/m,t_1)\right.\nonumber\\
 & &\left.\int d^3q|\mbox{\boldmath$k-q$}|p(\mbox{\boldmath$x$}+
\mbox{\boldmath$k$}t_1/m,\mbox{\boldmath$q$},t_1)\right\}.
 \label{meanfreetime}
\end{eqnarray}
Comparing this with the the identity
\[
t_\ell=\int_0^\infty dt\,e^{-t/t_\ell},\] we see that if the
integrand in the exponential in (\ref{meanfreetime}) had been
independent of $t_1$, then we could have identified $t_\ell$ as
\begin{equation}
t_\ell=\frac{m^2}{\sigma}\left\{\rho(\mbox{\boldmath$x$}) \int
d^3q|\mbox{\boldmath$k-q$}|p(\mbox{\boldmath$x,q$})\right\}^{-1}.
\label{meanfree2}
\end{equation}
Since $e^{-16}\approx 10^{-7}$, only values of $t$ less than
$16t_\ell$ contribute significantly to the integral $\int
...dt$ in (\ref{meanfreetime}). By the mean-value theorem, we may
write the argument of the exponential in (\ref{meanfreetime}) as
$-t/t_2$, where $t_2$ is the expression (\ref{meanfree2})
evaluated  at $\mbox{\boldmath$x$}+\mbox{\boldmath$k$}t_3/m$ for
the intermediate value $t_3<16t_\ell$. Then (given that $\partial
t_\ell=O(t_\ell)$) the correction to (\ref{meanfree2}) is
$O(t_\ell^2)$, and so (\ref{meanfree2}) can be taken as the mean
free time.
Now, $p=\overline{p}+O(t_\ell)$, so we may put $p=\overline{p}$ in 
(\ref{meanfree2}). Let  
\[R=
\int d^3q
|\mbox{\boldmath$k-q$}|\overline{p}(\mbox{\boldmath$x,q$})
/\overline{p}(\mbox{\boldmath$x,k$})\] In the ratio $R$, the
partition function cancels. The exponent in the Maxwellians is
\begin{eqnarray*}
-\frac{\beta}{2m}\mbox{\boldmath$q\cdot
q$}-\mbox{\boldmath$q\cdot\zeta$}+
\frac{\beta}{2m}\mbox{\boldmath$k\cdot k$}&+&\mbox{\boldmath$k\cdot\zeta$}=\\
-\frac{\beta}{2m}(\mbox{\boldmath$q-k$})\cdot(\mbox{\boldmath$q-k$})&-&
(\mbox{\boldmath$q-k$})\cdot((\beta/m)\mbox{\boldmath$k$}+
\mbox{\boldmath$\zeta$}).
\end{eqnarray*}
We change the variables of integration to {\boldmath$q-k$},
rewritten {\boldmath$q$}, to get
\[
R=\int
d^3q|\mbox{\boldmath$q$}|\exp\left\{-\frac{\beta\mbox{\boldmath$q\cdot
q$}}
{2m}-\mbox{\boldmath$q$}\cdot\left(\frac{\beta}{m}\mbox{\boldmath$k$}+
\mbox{\boldmath$\zeta$}\right)\right\}.\] Put
\begin{equation}
\mbox{\boldmath$\kappa$}=\left(\frac{m}{\beta}\right)^{1/2}\left(
\frac{\beta}{m}
\mbox{\boldmath$k$}+\mbox{\boldmath$\zeta$}\right)=c^{-1}\left(
\frac{\mbox{\boldmath$k$}}{m}-\mbox{\boldmath$u$}\right);
\label{peculiar}
\end{equation}
thus, $c${\boldmath$\kappa$} is the peculiar velocity \cite{Chapman}, p.
27. Let {\boldmath$q$}$^\prime=(\beta/m)^{1/2}${\boldmath$q$}.
Then
\[
|\mbox{\boldmath$q$}|d^3
q=(m/\beta)^2|\mbox{\boldmath$q$}^\prime|\,d^3q^\prime.\] Dropping
the dash, and choosing the $q_3$ axis along {\boldmath$\kappa$},
we get
\begin{eqnarray*}
R&=&\int
d^3q|\mbox{\boldmath$q$}|\frac{m^2}{\beta^2}\exp\left\{-\frac{1}{2}q^2
-\mbox{\boldmath$q\cdot\kappa$}\right\}\\
&=&\int_0^\infty q^3 dq\int_0^\pi
\sin\theta\,d\theta\int_0^{2\pi}d\varphi
\frac{m^2}{\beta^2}\exp\left\{-\frac{1}{2}q^2-q\kappa\cos\theta\right\}\\
&=&2\pi\int_0^\infty
q^3\,dq\frac{m^2}{\beta^2}e^{-q^2/2}(q\kappa)^{-1}\left(
e^{q\kappa}-e^{-q\kappa}\right).
\end{eqnarray*}
Thus,
\[ R=\frac{2\pi
m^2}{\beta^2\kappa}e^{\kappa^2/2}\left(I_2(-\kappa)-I_2(\kappa)\right)\]
where we use the functions $I_n$ \cite{Fisher}
\[ I_n(\kappa)=\int_0^\infty e^{-1/2(q+\kappa)^2}q^ndq.\]
Put
\[ F(\kappa)=\kappa\exp\{-\frac{1}{2}\kappa^2\}\left(I_2(-\kappa)-I_2
(\kappa)\right)^{-1}.\]
Then the mean free time is given
by
\begin{equation}
\overline{\rho}\overline{p}t_\ell=\frac{\beta^2}{2\pi\sigma}F(\kappa).
\label{meanfreetime3}
\end{equation}
\subsection{Galilean invariance}
We now show that the total transition probability {\em rate} is
invariant under ${\cal G}$. We have split up the time evolution into 
sub-processes,
in one of which particle $A$ at {\boldmath$x$} with momentum
{\boldmath$k$} has a free path of length $r=sa$, and then collides
with particle $B$ of momentum {\boldmath$q$} between $r$ and
$r+dr$. The rate of this process was taken to be the same as that
of a process in which $B$ was brought to rest by a change of
inertial frame, and makes a collision between $r^\prime$ and
$r^\prime+dr^\prime$, the positions as viewed by the observer
moving with  $B$. This is obviously necessary if the theory is to
be ${\cal G}$- invariant. We now show that it is also sufficient:
the {\em rate} of the one physical process, as viewed in two
relatively moving frames, will be shown to be the same. The key is
to remark that the time interval $dt$ in which $A$ collides with
$B$ after its free path is the same in all inertial frames, in
contrast to the distance gone, the free path $r$ and its increment
$dr$, which depend on the speed of $A$.

Suppose that ${\cal O},{\cal O}^\prime$ are two inertial
observers, with ${\cal O}^\prime$ moving with velocity
$-\mbox{\boldmath$v$}$ relative to ${\cal O}$, such that
$t^\prime=t$ and
\begin{equation}
\mbox{\boldmath$x$}^\prime=\mbox{\boldmath$x$}+\mbox{\boldmath$v$}t.
\end{equation}
The field variables as viewed by ${\cal O}^\prime$ are
$p^\prime,\;\;\rho^\prime=ma^{-3}N^\prime,\;\;
\mbox{\boldmath$k$}^\prime,\;\;\mbox{\boldmath$q$}^\prime$, where
\begin{eqnarray}
p^\prime(\mbox{\boldmath$x$}^\prime,\mbox{\boldmath$k$}^\prime,t)
&=&p(\mbox{\boldmath$x$},\mbox{\boldmath$k$},t)\nonumber\\
N^\prime(\mbox{\boldmath$x$}^\prime,\mbox{\boldmath$k$}^\prime,t)&=&N(
\mbox{\boldmath$x,k$},t), \label{Galrelation}
\end{eqnarray}
but along a path of a moving particle, the probability densities
$w$ and $w^\prime$ must satisfy
\begin{equation}
w(\mbox{\boldmath$x,k,q$},r)dr=w^\prime(\mbox{\boldmath$x$}^\prime,
\mbox{\boldmath$k$}^\prime,\mbox{\boldmath$q$}^\prime,r^\prime)dr^\prime.
\label{NSC}
\end{equation}
Here,
\[
\mbox{\boldmath$k$}^\prime=\mbox{\boldmath$k$}-m\mbox{\boldmath$v$}
\hspace{1in}\mbox{\boldmath$q$}^\prime
=\mbox{\boldmath$q$}-m\mbox{\boldmath$v$}.\] The point
{\boldmath$x+\hat{k}$}$r_1=\mbox{\boldmath$x+k$}t_1/m$ on the free
path, is assigned the coordinate
{\boldmath$x$}+{\boldmath$k$}$^\prime t_1/m$ by ${\cal O}^\prime$.
The righthand side of (\ref{NSC}) is calculated by ${\cal
O}^\prime$ using (\ref{collisionterm}) and (\ref{W}) to be
\begin{eqnarray*}
& &\exp\left\{-m^{-1}\sigma\int_{\mbox{\boldmath$x$}^\prime}
^{\mbox{\boldmath$x$}^\prime+ \mbox{\boldmath$\hat{k}$}^\prime
r^\prime}\rho^\prime(\mbox{\boldmath$x$}^\prime(t_1),t_1)
p^\prime(\mbox{\boldmath$x$}^\prime(t_1),
\mbox{\boldmath$q$}^\prime,t_1) |\mbox{\boldmath$k$}^\prime-
\mbox{\boldmath$q$}^\prime|\frac{dr_1^\prime}{|\mbox{\boldmath$k$}^\prime|}
\right\}\\
& &N^\prime(\mbox{\boldmath$x$}^\prime,0)p^\prime(
\mbox{\boldmath$x$}^\prime,\mbox{\boldmath$k$}^\prime,0)
\sigma\rho^\prime\left(\mbox{\boldmath$x$}^\prime+\mbox{\boldmath$\hat{k}$}
^\prime
r^\prime,t)\right)\,|\mbox{\boldmath$k$}^\prime-\mbox{\boldmath$q$}
^\prime|\,p^\prime\left(\mbox{\boldmath$x$}^\prime(t),
\mbox{\boldmath$q$}^\prime,t\right)\frac{dr^\prime}
{m|\mbox{\boldmath$k$}|^\prime}.
\end{eqnarray*}
Here, $r_1^\prime=|\mbox{\boldmath$k$}^\prime|t_1/m$. Then, by
using (\ref{Galrelation}) and the remark that
\[
\frac{dr_1^\prime}{|\mbox{\boldmath$k$}^\prime|}=\frac{dt_1}{m}
=\frac{dr_1}{|\mbox{\boldmath$k$}|},\hspace{.5in}0\leq r_1\leq
r,\] we see that ${\cal O}^\prime$ and ${\cal O}$ assign the same
probability to every event, so the integrals over {\boldmath$k,q$}
are also the same.

\input 02diff04

%% file: 02diff04.tex
\section{Compressible Navier-Stokes with Temperature}
\subsection{The Fundamental Relation}
The number of particles, thermalised plus unthermalised,
is conserved {\em locally}; that is, in any region, however small,
the loss in particles is the same as the integral of the current
over the boundary. This local conservation law does not apply to
the thermalised subset of particles. A particle can cease to be
thermalised at {\boldmath$x$} and collide at {\boldmath$y$},
thereby returning to the fold after a time in the unthermalised
state. We now show how the total probability $\mu$ is related to the
thermalised part, $\overline{\mu}$. At time $t_0$, any particle at
{\boldmath$x$} of momentum {\boldmath$k$} must have been from a
thermalised sample at some earlier time, $t_0-t$, at the point
$\mbox{\boldmath$x-k$}t/m$, and remained unthermalised at
{\boldmath$x$}, which it passes at time $t_0$. The probability of
thermalising exactly at {\boldmath$x$} is zero. It must thermalise
at some later time, say after it has travelled for a free time
$\tau=t^\prime$. Then $t^\prime>t$ must hold. We first compute the
probability arising from a hop of fixed size
$r^\prime=|\mbox{\boldmath$k$}| t^\prime/m$. The rate at which
this occurs is
\[
(1/t^\prime)w(\mbox{\boldmath$x-k$}t/m,\mbox{\boldmath$k$},t_0-t,t^\prime).\]
In the interval of time from $t$ to $t+dt$, the number of hops is
$\mbox{\em rate}\times dt$, so the probability of a particle being
at {\boldmath$x$} at time $t_0$ with momentum {\boldmath$k$}, and
having free time $t^\prime$, is, at time $t_0$,
\begin{eqnarray*}
P(t^\prime)&:=&\mbox{Prob}_\mu\left\{\omega: {\cal
N}_{\mbox{\boldmath$x$}} (\omega)=1\cap\mbox{\boldmath${\cal
P}_x$}=\mbox{\boldmath$k$}|\tau=t^
\prime\right\}=\\
&=& \int_0^{t^\prime}dt (t^\prime)^{-1}\overline{N}
(\mbox{\boldmath$x-k$}t/m,t_0-t)\overline{p}
(\mbox{\boldmath$x-k$}t/m,\mbox{\boldmath$k$},t_0-t)\\
& &w(\mbox{\boldmath$x-k$}t/m,\mbox{\boldmath$k$},t_0-t,t^\prime).
\end{eqnarray*}
Now, $w$ is a density of probability (of collision) as a function
of free path size $t^\prime$, so the total contribution to $\mu$ due
to hops from one side of {\boldmath$x$} to the other along the
line of {\boldmath$k$} is
\begin{eqnarray}
N(\mbox{\boldmath$x$},t_0)p(\mbox{\boldmath$x,k$},t_0)&=&\int_0^\infty
P(t^\prime)dt^\prime\nonumber\\
&=&\int_0^\infty
\frac{dt^\prime}{t^\prime}\int_0^{t^\prime}dt\,\overline{N}
(\mbox{\boldmath$x-k$}t/m,t_0-t)\overline{p}
(\mbox{\boldmath$x-k$}t/m,\mbox{\boldmath$k$},t_0-t)\nonumber\\
& &w(\mbox{\boldmath$x-k$}t/m,\mbox{\boldmath$k$},t_0-t,t^\prime).
\label{thenewnavier}
\end{eqnarray}
This is the fundamental relation.
The logarithmic divergence at $t^\prime=0$ is only apparent, if
the functions entering the integral are smooth enough. If so we
can expand in Taylor series in $t^\prime$ up to $O(t_\ell)$ around
the point $\mbox{\boldmath$x,k$},t_0$, at which the functions are
evaluated:
\begin{eqnarray}
Np&=&\int_0^\infty
\frac{dt^\prime}{t^\prime}\int_0^{t^\prime}dt\left\{\overline{N}
\overline{p}w(t^\prime)-t(\mbox{\boldmath$k\cdot\partial$}/m+\partial_0)
\left(\overline{N}\overline{p}w(t^\prime)\right)\right\}\nonumber\\
&=&\overline{N}\overline{p}-\frac{1}{2}\left(\frac{\mbox{\boldmath$k\cdot
\partial$}}{m}+\partial_0\right)\overline{N}\overline{p}t_\ell
\label{basic}
\end{eqnarray}
because of (\ref{sumisone2}) and (\ref{relax}). Here, $\partial_0$
means $\partial/\partial t_0$. This is a ${\cal G}$-invariant
splitting; for, the equation shows that $Np$ and
$\overline{N}\overline{p}$ differ by a quantity of order $t_\ell$
in smallness, so $\overline{N}\overline{p}$ transforms correctly
up to first order. But $t_\ell$ is ${\cal G}$-invariant, and
{\boldmath$k\cdot
\partial$}$/m+\partial_0$ is a ${\cal G}$-invariant operator (on
fields that transform correctly), so $\overline{N}\overline{p}$
transforms correctly up to $O(t_\ell^2)$; and so on.

The second term in (\ref{basic}) is responsible for the
dissipation. Putting in (\ref{meanfreetime3}) for $t_\ell$, we see
that the
density cancels; so the conductivity and viscosity
of a gas are independent of the density. This is Maxwell's
famous result.

Taking the
expectation values of $\chi={\cal N}, \mbox{\boldmath${\cal P}$}$
or ${\cal E}$ gives us the relation between the mean of the
thermalised particles and the true means of all the particles. In
this, we need to evaluate $\partial_0$ applied to the thermalised
variables. Here we can assume that these obey Euler's equations,
since these hold up to first order, and the operator $\partial_0$
acts only on small quantities. The integral over $w$ in
(\ref{thenewnavier}) acts as a smoothing operator, so we expect
$\mu$ to be more regular than $\overline{\mu}$.

\subsection{The Euler equations}
The current of the conserved variable $\chi$ is
\begin{equation}
\mbox{\boldmath$j$}_\chi:=\int
d^3k\,Np\mbox{\boldmath$\Upsilon$}\chi,\hspace{.5in}\mbox
{where \boldmath$\Upsilon$}=\mbox{\boldmath${\cal P}$}/m.
\end{equation}
This gives us the dynamics
\begin{equation}
\frac{d\langle\chi\rangle}{dt}=-\partial_j\int d^3k\chi
(\mbox{\boldmath$k$})N(\mbox{\boldmath$x$})p(\mbox{\boldmath$x,k$})
\frac{k_j}{m}=-\partial_j\langle\chi\frac{k_j}{m}\rangle.
\label{conservedcurrent}
\end{equation}
These equations, for $\chi$ running over the slow variables $m$,
{\boldmath$k$} and {\boldmath$k\cdot k$}$/2$, can replace {\em
C-N-S-T} for dilute inert gases. Putting $\chi=m$ gives the usual
`equation of continuity'
\begin{equation}
\frac{d\rho}{dt}+\partial_j(u_j\rho)=0, \label{masscons}
\end{equation}
which is exact.
We shall show how to compute the other equations up to order
$t_\ell$ for our choice of $C$. Our strategy is to
use (\ref{basic}) in (\ref{conservedcurrent}), allowing us to
evaluate the righthand side in terms of the means in
$\overline{\mu}$; we then use (\ref{basic}) again to rewrite this
in terms of the true means.

The zero$^{\rm th}$ order approximation to (\ref{basic}), namely
$Np=\overline{N}\overline{p}$, can be put in (\ref{conservedcurrent}) and
computed exactly: we get the Euler equations; this is shown very smoothly
by using the cumulant generating function,  $\log Z$.

For the momentum, put $\chi={\cal P}_{\mbox{\boldmath$x$}}^i$. Then we have for each
{\boldmath$x$},
\begin{eqnarray*}
\frac{\partial(Nu^i)}{\partial t}&=&-\partial_j\left(N{\bf E}_{\overline{p}}
[\Upsilon^i\Upsilon^j]\right)\\
&=&-\partial_j\left(N(\langle\Upsilon^i\Upsilon^j\rangle_T+u^iu^j)\right)
\end{eqnarray*}
Since
\[
\langle{\cal P}^i{\cal P}^j\rangle_T=\frac{\partial^2\log Z}
{\partial\zeta_i\partial\zeta_j}=mk_{_B}\Theta\delta_{ij},\]
we get the Euler equation for momentum conservation:
\begin{equation}
\frac{\partial}{\partial t}(\rho u^i)+\partial_j(\rho u^iu^j)+
\partial_i(\rho k_{_B}\Theta/m)=0.\label{momentumcons}
\end{equation}
Finally, for the energy, put $\chi={\cal E}_{\mbox{\boldmath$x$}}
=\mbox{\boldmath$k\cdot k$}/(2m)$ in (\ref{conservedcurrent}), which then
becomes
\begin{equation}
\dot{E}=-\mbox{div}\,\left(m^{-1}N{\bf E}_{\overline{p}}
[{\cal E}\mbox{\boldmath${\cal P}$}]\right).
\end{equation}
Now, for each {\boldmath$x$},
\begin{eqnarray*}
{\bf E}_{\overline{p}}[{\cal E}{\cal P}^j]&=&\langle{\cal E}{\cal P}^j
\rangle_T+\langle{\cal E}\rangle\langle{\cal P}^j\rangle\\
&=&\frac{\partial^2\log Z}{\partial\beta\partial\zeta_j}+NEmu^j\\
&=&-\frac{m\zeta_j}{\beta^2}+mNu^j\left(k_{_B}\Theta+\frac{1}{2}
\mbox{\boldmath$u\cdot u$}\right)
\end{eqnarray*}
from (\ref{Z}). Since
\[
\zeta^j=-\beta u^j\hspace{.2in} \mbox{ and
}\hspace{.2in}E=mN(3k_{_B}\Theta/m+ \mbox{\boldmath$u\cdot
u$})/2\]
we can collect terms to get the Euler equation for energy
conservation:
\begin{equation}
\frac{\partial}{\partial t}
\left(\rho(3k_{_B}\Theta/m+\mbox{\boldmath$u\cdot u$})\right)/2
+\mbox{div}\,\left(\rho\mbox{\boldmath$u$}(5k_{_B}\Theta/m+
\mbox{\boldmath$u\cdot u$})/2\right)=0.
\label{energycons}
\end{equation}
The pressure appearing in the usual form of Euler's equations is
here replaced by $\rho k_{_B}\Theta/m$, the pressure for a perfect
gas. This differs from the static pressure (\ref{pressure}) by
terms which vanish in the low density limit. We shall use the
Euler equations, which are first-order PDE in space and time, to
relate $\partial_0$ to a first-order gradient. Let $D:=
\mbox{\boldmath$u\cdot\partial$}+\partial_0$ be the Lagrange
material derivative. Then we have
\begin{lemma}
The Euler equations (\ref{masscons}), (\ref{momentumcons}) and
(\ref{energycons}) imply the short Eulers:
\begin{eqnarray}
D\rho&=&-\rho\partial_ju^j\\
Du^i&=&-k_{_B}\rho^{-1}\partial_i\left(\rho\Theta/m\right)\\
D\Theta&=&-\frac{2}{3}\Theta\partial_ju^j.
\end{eqnarray}
\label{shorteuler}
\end{lemma}
Proof. `It does not seem necessary to reproduce the details of
this proof; the mathematician will be able to construct them for
himself, while the physicist will probably not wish to be detained
over
them' \cite{Jeans}.\\
The ${\cal G}$-invariance of this form is obvious.

\subsection{Calculations}
It is the full state $\mu$, rather than the thermalised part,
$\overline{\mu}$, that is usually measured in experiments. For example, a
measurement of density can be made by noting the absorption of a laser
passing through the gas. The scattering of photons with
particles makes no distinction between thermalised and
non-thermalised particles. Again, one can measure the temperature
by a probing thermometer, which would tend to thermalise any particles
that struck it, whether they were thermalised before or not.
However, the means in the thermalised state are much easier to
calculate; the state $\overline{p}$ is Gaussian, and the fields
are independent at different points (at the same time, say $t_0$).
This enables us to relate the extensive to the intensive
variables. For $\mu$, the intensive variables have not even been
defined yet.

In the Boltzmann equation, authors write the phase-space density as
a product $N(\mbox{\boldmath$x$})f(\mbox{\boldmath$x,k$})$ with some
hesitation, as ${\cal N}$ and {\boldmath${\cal P}$} are not independent
random variables, even in the Maxwellian. Not to worry. For the general
state $\mu$ we define
\begin{equation}
N(\mbox{\boldmath$x$})=\mbox{Prob}_\mu\left\{\omega:{\cal N}
(\mbox{\boldmath$x$})=1\right\},
\end{equation}
and $p(\mbox{\boldmath$x,k$})$ is the conditional probability
\begin{equation}
p(\mbox{\boldmath$x,k$})=\mbox{Prob}_\mu\left\{\mbox{\boldmath${\cal P}_x=k$}
|{\cal N}(\mbox{\boldmath$x$})=1\right\}.
\end{equation}
By Bayes's definition,
\begin{equation}
\mu_{\mbox{\boldmath$x$}}(\omega_{\mbox{\boldmath$x$}})=N(\mbox{\boldmath$x$})
p(\mbox{\boldmath$x,k$}).
\end{equation}
Then we may define $\rho=mN/a^3$. The momentum density also has a
definition in terms of $\mu$, which does not assume that $p$ is Maxwellian, by
\[
\mbox{\boldmath$\varpi$}=\mbox{\boldmath$\Pi$}/a^3=
{\bf E}_\mu[\mbox{\boldmath${\cal P}$}]/a^3.\]
We can then define the velocity field, without
recourse to information geometry, by
\[
\mbox{\boldmath$u$}=\mbox{\boldmath$\varpi$}/\rho,\] provided that
the density never vanishes. This is indeed so, as we see from the
basic equation (\ref{thenewnavier}). Finally, we define the {\em
thermal energy per unit mass}, $e$, to be given in terms of a
`temperature' $\Theta$ at each {\boldmath$x$} by
\begin{equation}
e=\frac{3}{2m}k_{_B}\Theta:=m^{-1}{\bf E}_p[{\cal
E}_{\mbox{\boldmath$x$}}]-\frac{1}{2}\mbox{\boldmath$u\cdot u$}.
\end{equation}
This definition of temperature is ${\cal G}$-invariant, as it can
also be written as $mc^2{\bf E}_p[\kappa_i\kappa_i)/2]$.

Suppose that we know the fields $\rho,\Theta$ and {\boldmath$u$},
at time=$t_0$, and thus also their space gradients. We can use
(\ref{basic}) and the short Eulers to find the barred parameters
of the Maxwellian $\overline{p}$, to first order in $t_\ell$. When
any expression is multiplied by $t_\ell$, we are able to replace
any thermalised parameters by the above unbarred parameters, or
{\em vice versa}, with only a second order error. So the analysis
reduces to linear algebra.

For any local random variable $\chi$, slow or not, $a^{-3}\chi$ is
its density, and
\[
a^{-3}\langle\chi\rangle:=\int d^3k\,m^{-1}\rho(\mbox{\boldmath$x$})
p(\mbox{\boldmath$x,k$})\chi(\mbox{\boldmath$k$})\hspace{.4in}a^{-3}\langle
\overline{\chi}\rangle:=\int
d^3k\,m^{-1}\overline{\rho}(\mbox{\boldmath$x$})
\overline{p}(\mbox{\boldmath$x,k$})\chi(\mbox{\boldmath$k$}).\]
To help in the evaluation of various derivatives of integrals that
occur here, we note that while $F(\kappa)$ is a complicated
function of {\boldmath$u$} and $\beta$, the integrals arising can
be evaluated if we change the variable of integration from
{\boldmath$k$} to {\boldmath$\kappa$}, given in (\ref{peculiar}).
We do this for each {\boldmath$x$}, and it is valid provided that
we keep the derivatives {\boldmath$\partial$} and $\partial_0$ to
the left of the expression. We note that
$d^3k=(m/\beta)^{3/2}\,d^3\kappa$.
Then using (\ref{basic}) and (\ref{meanfreetime3}) we have to order
$t_\ell$:
\begin{eqnarray}
a^{-3}(\langle\chi\rangle-\langle\overline{\chi}\rangle)&=&-
\frac{1}{4\pi\sigma m}\int d^3k
\left(\frac{\mbox{\boldmath$k\cdot\partial$}}{m}+\partial_0\right)
F(\kappa)\beta^2\chi(\mbox{\boldmath$k$})\nonumber\\
&=&-\frac{1}{4\pi\sigma}\left(\frac{m}{k_{_B}}
\right)^{1/2}\partial_j\Theta^{-1/2}\int d^3\kappa(c\kappa_j+u_j)
\chi[m(c\mbox
{\boldmath$\kappa$}+\mbox{\boldmath$u$})]F(\kappa)\nonumber\\
& &-\frac{1}{4\pi\sigma}\left(\frac{m}{k_{_B}}\right)^{1/2}
\partial_0\Theta^{-1/2}\int d^3\kappa\;
\chi[m(c\mbox{\boldmath$\kappa+u$})]F(\kappa). \label{barrelation}
\end{eqnarray}
Let us put
\begin{equation}
\lambda_n:=\frac{m}{\sigma}\left(\frac{m}{k_{_B}}\right)^{1/2}\int_0^
\infty\kappa^{2n} F(\kappa)d\kappa,\hspace{.5in}n=1,2,3.
\end{equation}
On putting $\chi=m$, the term odd in {\boldmath$\kappa$} is zero,
and we get
\begin{equation}
\rho=\overline{\rho}-\lambda_1\partial_j\left(\Theta^{-1/2}u_j\right)
-\lambda_1\partial_0\Theta^{-1/2}.
\label{barrho}
\end{equation}
We now put $\chi=${\boldmath$k$}, and
$\mbox{\boldmath$\hat{\kappa}$}=(\sin\theta\cos\varphi,
\sin\theta\sin\varphi,\cos\varphi)$. In the following calculations, we
use that
\begin{equation}
\int d\Omega \hat{\kappa}_i\hat{\kappa}_j= \int
\sin\theta\,d\theta\,d\varphi\,
\hat{\kappa}_i\hat{\kappa}_j=\frac{4\pi}{3}\delta_{ij},
\end{equation}
and that the integral of odd powers of $\hat{\kappa}$ are zero.
Then (\ref{barrelation}) gives
\begin{equation}
\varpi_i-\overline{\varpi}_i=-\frac{1}{3}\frac{k_{_B}}{m}
\lambda_2\partial_i\Theta^{1/2}-\lambda_1\partial_j\left(
\Theta^{-1/2}u_iu_j\right)-\lambda_1\partial_0\left(\Theta^{-1/2}u_i\right).
\label{barmom}
\end{equation}
We now put $\chi=\mbox{\boldmath$k\cdot k$}/(2m)$; we get
\begin{eqnarray*}
a^{-3}(E-\overline{E})&=&-\frac{m}{8\pi\sigma}\left(\frac{m}{k_{_B}}
\right)^{1/2}\partial_0\Theta^{-1/2}\int F(\kappa)\kappa^2d\kappa d\Omega
\left(c\kappa_j+u_j\right)\left(c\kappa_j+u_j\right)\\
&-&\frac{m}{8\pi\sigma}\left(\frac{m}{k_{_B}}\right)^{1/2}
\partial_j\Theta^{-1/2}
\int \kappa^2d\kappa d\Omega\,F(\kappa)\left(c\kappa_j+u_j\right)
\left(c\kappa_\ell+u_\ell\right)\left(c\kappa_\ell+u_\ell\right),
\end{eqnarray*}
which simplifies to
\begin{eqnarray}
a^{-3}(E-\overline{E})&=&-\frac{5k_{_B}\lambda_2}{6m}\partial_j\left(
\Theta^{1/2}u_j\right)-\frac{\lambda_1}{2}\partial_j
\left(\Theta^{-1/2}u_ju_\ell u_\ell\right)\nonumber\\
& &-\frac{k_{_B}\lambda_2}{2m}\partial_0\Theta^{1/2}-
\frac{\lambda_1}{2}\partial_0\left(\Theta^{-1/2}u_\ell
u_\ell\right). \label{barenergy}
\end{eqnarray}
We can use these results to relate $\langle\chi\rangle$ and
$\langle\overline{\chi}\rangle$ for any polynomial.
Let
\begin{equation}
\delta\chi=\langle\chi\rangle-\langle\overline{\chi}\rangle.
\end{equation}
Then up to $O(t_\ell)$, $\delta$ is a derivation, and we have for example,
\begin{equation}
\delta(\rho\mbox{\boldmath$u\cdot u$})=2u_j\delta(\rho u_j)-
\mbox{\boldmath$u\cdot u$}\delta\rho.
\end{equation}
\begin{lemma}
We have the relation
\begin{equation}
\delta(\rho\Theta)=-\frac{5}{9}\lambda_2\Theta^{1/2}\partial_ju_j-
\frac{1}{3}m\lambda_2D\Theta^{1/2}=-\frac{4\lambda_2}{9}\Theta^{1/2}
\partial_ju_j. \label{ginvarianttemp}
\end{equation}
\label{short}
\end{lemma}
Proof.
\begin{eqnarray*}
\frac{3}{2}k_{_B}\delta(\rho\Theta/m)&=&\frac{3}{2}\frac{k_{_B}}{m}
(\rho\Theta-\overline{\rho}\overline{\Theta})\\
&=&a^{-3}\delta E-u_i\delta\varpi_i+u_iu_i\delta\rho/2\\
&=&-\frac{5k_{_B}}{6m}\lambda_2\Theta^{1/2}\partial_ju_j-\frac{k_{_B}}{2m}
\lambda_2D\Theta^{1/2},
\end{eqnarray*}
making use of (\ref{barrho}), (\ref{barmom}) and
(\ref{barenergy}). This gives the result, using Lemma~(\ref{shorteuler})
to get the second form.

\subsection{Equations of Motion}
The equations of motion express the rate of change of the
slow variables (the conserved quantities) in terms of the mean of their
microscopic currents; we use (\ref{conservedcurrent}).

\subsubsection{Viscosity Terms}
Let $J_{\varpi_i}^j$ be the $j^{\rm th}$ component of the mean
{\em BBGKY} current density of the $i^{\rm th}$ component of
momentum in the state $\mu$, and let $J_{\varpi_i}^{\circ j}$ be
the Euler term $\rho u_iu_j+ k_{_B}\delta_{ij}\rho\Theta/m$. Let
$\overline{J}_{\varpi_i}^j$ be the same object averaged in the
state $\overline{\mu}$; as found in (\ref{momentumcons}), this is
the Euler term with the barred values of the fields. Then the
choice $\chi=k_i$ in (\ref{conservedcurrent}) gives, using
(\ref{barrelation}),
\begin{eqnarray*}
J_{\varpi_i}^j&=&\int d^3k\frac{k_ik_j}{m^2}\rho p\\
&=&\overline{J}_{\varpi_i}^j-\frac{m}{4\pi\sigma}
\left(\frac{m}{k_{_B}}\right)^{1/2}\partial_0\Theta^{-1/2}\int
\kappa^2F(\kappa)d\kappa\,d\Omega(c\kappa_i+u_i)(c\kappa_j+u_j)\\
&-&\frac{m}{4\pi\sigma}\left(\frac{m}{k_{_B}}\right)^{1/2}\partial_\ell
\Theta^{-1/2}\int\kappa^2F(\kappa)d\kappa\,(c\kappa_i+u_i)(c\kappa_j+u_j)
(c\kappa_\ell+u_\ell).
\end{eqnarray*}
Noting that the odd powers of $\kappa_i$ integrate to zero, this reduces to
\begin{eqnarray}
J_{\varpi_i}^j&=&\overline{J}_{\varpi_i}^j-\frac{k_{_B}\lambda_2}{3m}
\left[\partial_\ell\left(\Theta^{1/2}u_\ell\right)\delta_{ij}+
\partial_j\left(\Theta^{1/2}u_i\right)+\partial_i
\left(\Theta^{1/2}u_j\right)\right]\nonumber\\
&-&\lambda_1\left[\partial_\ell\left(\Theta^{-1/2}u_iu_ju_\ell\right)+
\partial_0\left(\Theta^{-1/2}u_iu_j\right)\right]-\frac{k_{_B}\lambda_2}
{3m}\partial_0\Theta^{1/2}\delta_{ij}.
\label{nearlythere}
\end{eqnarray}
Note that derivatives act on all functions to their right. We get
our momentum equation by relating $\overline{J}_{\varpi_i}^j$ to
$J_{\varpi_i}^{\circ j}$:
\begin{equation}
\overline{J}_{\varpi_i}^j=J_{\varpi_i}^{\circ j}
-u_j\delta(\varpi_i)
-u_i\delta(\varpi_j)+\left(u_iu_j\delta\rho\right)-\frac{k_{_B}}{m}
\delta(\rho\Theta)\delta_{ij},
\end{equation}
for which we use (\ref{barrho}), (\ref{barmom}) and
Lemma~\ref{short}. Putting this in (\ref{nearlythere}), we see
that most terms cancel, leaving
\begin{equation}
J_{\varpi_i}^j=J_{\varpi_i}^{\circ j}-\frac{1}{3}\frac{k_{_B}}{m}
\lambda_2\Theta^{1/2}\left(\partial_ju_i+\partial_iu_j-\frac{2}{3}
\partial_\ell u_\ell\;\delta_{ij}\right).
\label{momentum}
\end{equation}
The kinetic pressure is traceless, so there is no bulk viscosity
\cite{Stokes}; we get a very special case of the general
equations, at the edge of possible values. It could be that the
Stokes relation only occurs when, as here, there is no interaction
whatsoever between the particles outside the hard core. It is
interesting that $\lambda_1$ does not occur in the answer, and
that all the terms in $\partial_0$ cancel out without recourse to
the short Euler equations, Lemma~(\ref{shorteuler}). The viscosity
coefficient increases as $\Theta^{1/2}$ with temperature, like
Enskog's and Chapman's prediction from Boltzmann's equation.

\subsubsection{Equation of motion for the energy}
Let $J_{_E}^j$ be the mean energy current density, with a bar if
the state $\overline{\mu}$ is used, and let
\begin{equation}
J_{_E}^{\circ j}:=\rho
u_j\left(\frac{5}{2}\frac{k_{_B}}{m}\Theta+\frac{1}{2}u_iu_i\right)
\end{equation}
be the Euler energy current. Putting $\chi=k_ik_i/(2m)$ in
(\ref{conservedcurrent}) gives
\begin{equation}
J_{_E}^j=\overline{J}_{_E}^j+\partial_0[A]+\partial_\ell[B],
\end{equation}
where
\begin{eqnarray}
A&=&-\frac{m\Theta^{-1/2}}{8\pi\sigma}\left(\frac{m}{k_{_B}}\right)^{1/2}
\int \kappa^2\,d\kappa\,d\Omega
F(\kappa)\left(c\kappa_i+u_i\right)\left(c\kappa_i+u_i\right)
\left(c\kappa_j+u_j\right)\nonumber\\
&=&-\frac{\lambda_1}{2}\Theta^{-1/2}u_iu_iu_j-\frac{5k_{_B}\lambda_2}{6m}
\Theta^{1/2}u_j, \label{A}
\end{eqnarray}
\begin{eqnarray}
B&=&-\frac{m\Theta^{-1/2}}{8\pi\sigma}\left(\frac{m}{k_{_B}}\right)^{1/2}
\int \kappa^2\,d\kappa\,d\Omega
F(\kappa)\left(c\kappa_\ell+u_\ell\right)\left(c\kappa_i+u_i\right)
\left(c\kappa_i+u_i\right)\left(c\kappa_j+u_j\right)\nonumber\\
&=-&\frac{k_{_B}^2\lambda_3}{6m^2}\delta_{j\ell}\Theta^{3/2}-\frac{7k_{_B}
\lambda_2}{6m}\Theta^{1/2}u_ju_\ell
-\frac{k_{_B}\lambda_2}{6m}\delta_{j\ell} \Theta^{1/2}u_iu_i
-\frac{\lambda_1}{2}\Theta^{-1/2}u_\ell u_iu_iu_j.\label{B}
\end{eqnarray}
Now, $\overline{J}_{_E}^j-J_{_E}^{\circ j}=-\delta J_{_E}^j$ has
six terms:
\begin{equation}
\delta J_{_E}^j=
\left[\frac{5k_{_B}}{2m}\left(\Theta\delta
\varpi_j+u_j\delta(\rho\Theta)-\Theta u_j\delta\rho\right)\right]
+\left[\frac{1}{2}u_iu_i\delta\varpi_j+u_iu_j\delta\varpi_i-
u_iu_iu_j\delta\rho\right].
\label{delta}
\end{equation}
This can be evaluated using (\ref{barrho}), (\ref{barmom}) and
Lemma~(\ref{short}). Collecting up $A$, $B$, and $-\delta
J_{_E}^j$, using the same method of proof as in
Lemma~(\ref{shorteuler}), the heat equation reduces to
\begin{eqnarray}
a^{-3}\frac{dE}{dt}&=&-\partial_jJ_{_E}^{\circ j}+
\frac{k_{_B}^2}{m}\left(\frac{\lambda_3}{4}-
\frac{5\lambda_2}{4}+\frac{5\lambda_1}{2}\right)\partial_j
\left[\Theta^{1/2}\partial_j\Theta\right]\nonumber\\
&+&\frac{5k_{_B}^2}{2m}\left(\lambda_1-\frac{\lambda_2}{3}\right)
\partial_j\left[\Theta^{3/2}\partial_j\log\rho\right]
-\frac{2k_{_B}}{9}\lambda_2\partial_j\left[\Theta^{1/2}u_j\partial_i
u_i\right]\nonumber\\
&+&\frac{k_{_B}}{3}\lambda_2\partial_j\left[\Theta^{1/2}u_i
\partial_iu_j\right]+\frac{k_{_B}}{3}\lambda_2\partial_j\left[\Theta^{1/2}
\partial_ju_iu_i/2\right].
\end{eqnarray}
The coefficient of the Fourier term is positive, since
\[
\frac{5}{2}-\frac{5}{4}\kappa^2+\frac{1}{4}\kappa^4>0.\] The new
term involves the logarithmic derivative of the density, whose
presence in a gas of a single component is denied in the
literature. It means that a gradient in the density contributes to
the heat current. This is the Dufour effect, also called the
diffusive thermal effect. The sign of the term does not need to be
definite. Some authors invoke Onsager symmetry to eliminate this
term without having to evaluate it, since its Onsager dual, the
Soret effect in the continuity equation for the mass, is absent. However
Onsager duality is not true here, because the state $p$ is
not in {\em LTE} and the transition rate depends on {\boldmath$x$}.
The present work does suggest
that the effect should be looked for experimentally, in say Helium, but
this is quite delicate since the Dufour effect is transitory, and becomes
less pronounced, and is masked by heat conduction and convection,
as time goes by.
The other terms have appeared in the literature \cite{Chen}. One can
check that the system of equations is
invariant under ${\cal G}$: `in the tradition of British
applied mathematics, it is not considered gentlemanly to press a
colleague for a proof' (G. Pistone).

\input 02diff05.tex

%% file: 02diff05.tex
\section{Conclusions}
We have shown that the `method of Maxwell' \cite{Maxwell} can be
made ${\cal G}$-invariant, and gives {\em C-N-S-T} with a Dufour
term.
The fluid equations we get are the following:
\begin{eqnarray}
\hspace*{-.2in}\frac{d\rho}{dt}&=&-\partial_j(\rho u_j)\\
\frac{d\varpi_i}{dt}&=&-\partial_j(\rho
u_iu_j)-\partial_iP+\lambda\partial_j\left[\Theta^{1/2}\left(\partial_ju_i+
\partial_iu_j-\frac{2}{3}\partial_\ell u_\ell\;\delta_{ij}\right)\right]\\
\frac{d}{dt}[\rho(e+u_iu_i/2)]&=&-\partial_j\left(\rho
u_j(e+u_iu_i/2)+u_jP\right)+\lambda_4\partial_j
\left(\Theta^{1/2}\partial_j\Theta\right)+
\lambda_5\partial_j\left(\Theta^{3/2}\partial_j\log\rho
\right)\nonumber\\
&+&\lambda\partial_j\left[\Theta^{1/2}u_i\left(\partial_ju_i+
\partial_iu_j-\frac{2}{3}\partial_\ell u_\ell\;\delta_{ij}\right)\right].
\label{theendresult}
\end{eqnarray}
The transport coefficients are independent of density, as found by
Maxwell. Our starting point is not the Boltzmann equation, but a
non-local integral equation, (\ref{thenewnavier}). The presence of
coefficient $\lambda_5$ is at variance with the results of Chapman
and Cowling \cite{Chapman}. The Boltzmann equation suffers from
the Hilbert paradox \cite{Balian2}, II, p. 348. Namely, the state
in Boltzmann's equation is parametrized by the initial
distribution $f$, which is a general integrable function of six
variables, whereas the hydrodynamic solutions are parametrized by
five functions of three variables; the set of hydrodynamic
solutions cannot describe the most general solution. To show that
they nevertheless provide a good approximate solution is an extra
burden if one uses the Boltzmann equation as the starting point.
In \cite{Chapman} this question is discussed but not solved on p
120; the authors refer the reader to \cite{Grad}. Chapman himself
has said that reading his book is like `chewing glass'
\cite{brush}. The fact is that the Boltzmann distribution $f$ is
too detailed a description for an easy move to thermodynamical
variables. Hilbert's paradox also shows up as follows. If the
initial state happens to be in {\em LTE}, then the collision term
in {\em BE} is zero, and (at that instant) the thermodynamic
variables follow the Euler equation, and the instantaneous rate of
entropy production is zero. However, fluids following the
equations supposedly derived from the {\em BE} do not at any
instant follow the Euler equations, or possess a zero rate of
entropy production, except in the special states with $\Theta$ and
{\boldmath$u$} independent of {\boldmath$x$}. This paradox lasts a
very brief time, after which the stirring due to the Euler
convection spoils {\em LTE}. For consistency between the {\em BE}
and the fluid equations, we are not allowed to choose an arbitrary
initial state for the {\em BE}; the small deviations from {\em
LTE} must be related to the fluxes of the theory \cite{Balescu},
p. 160. Our approach avoids the paradox: an initial state $\mu$
can be {\em LTE}, and then we would modify (\ref{thenewnavier}) in
the obvious way.

The point of view of the present paper differs from the usual one,
such as \cite{Boon}. We assume that just {\em after} a collision,
which we prefer to call a thermalisation, the particle is well
described as being in {\em LTE}, and almost independent of its
neighbours. This is a good time `to carve Nature at it joints'
\cite{Auyang}, p 341. It is during the free propagation that the
state loses its {\em LTE} property, since then particles from
regions of different density and temperature come together. They
are likely to be independent because `they have a different
history' \cite{Spohn}. Concerning the other phase of the dynamics,
our collision term involves the density at different space-time
points, whereas the Boltzmann kernel is local. In the models in
\cite{Boon}, several collisions are needed before a particle is
close to thermalised; the authors do, however, show a preference
for models, called `efficient', in which the collision output is
randomised over the available channels, and so thermalises
rapidly. In the present model, a particle thermalised after one
collision. Our result shows that the details of how many
collisions are needed does not affect the qualitative results,
though it changes the relationship between geometric cross-section
and mean free path. It is difficult to believe that the
simplification made here is responsible for the Dufour effect.

Our equations (\ref{theendresult}) may be less singular than {\em C-N-S-T}
since the presence of the Dufour term means that the symmetrised part of
the operator $\partial_0$ has a principal symbol of full rank, at least in
general position. It might be that {\em C-N-S-T} has no global smooth
solutions, or, even if it has some, it might be too hard for anyone to
prove it. Whatever the case, it might be easier to show that
(\ref{thenewnavier}) has smooth global solutions

We can generalise in various ways. If there is an external potential,
$\Phi$, it does not affect the
local state, because it cancels out in $\bar{\mu}$; however, it does
affect the hopping rates, and thus appears in the equations of
motion. In a paper \cite{Grasselli} we find the
equations of motion for a fluid moving in a potential, in a
non-galilean model. The same method can be applied to the present
model. It is possible to extend both models to the case of
inter-particle potentials by following a suggestion of Biler and
collaborators
\cite{Biler1,Biler2,Biler3,Biler4,Biler5,Biler6,Nad}. This gives a
macroscopic dynamics in which the rate of change of energy at a
point {\boldmath$x$} is governed by the mean field of all the
other particles. It seems unlikely that making the model more realistic
by including interactions will exactly cancel out the prediction of
the Dufour effect; thus this should be looked for in Helium or Argon.

 \vspace{.1in} \noindent{\bf Acknowledgements}. I thank P. Zweifel for
encouragement. This theory was developed at the Institute of
Physics of Sao Paulo, Brazil. I thank the University of Sao Paulo
for its generous support, and  Walter Wreszinski for arranging the
visit and for critical discussions. The work was completed at
MaPhySto, Aarhus; I owe this visit to Ole Barndorff-Nielsen, and
Goran Peskir. I am endebted to Roger Balian for an e-mail
correspondence, which led to the present version, which has fewer
serious errors than the original.